January 9, 2024

# Clinical Applications of Plantar Pressure Measurement

Kelsey Detels[1]

David Shin[1]

Harrison Wilson[1]

Shanni Zhou[1]

Andrew Chen[1]

Jessica Rosendorf[1]

Atta Taseh[1]

Bardiya Akhbari, Ph.D.[1]

Joseph H. Schwab, M.D., M.Sc.[1,2]

Hamid Ghaednia, Ph.D.[1,2]

**Affiliations:**

[1] Department of Orthopaedic Surgery, Massachusetts General Hospital, Harvard Medical School, Boston, MA 02114

[2] Center for Surgical Innovation and Engineering, Department of Orthopaedic Surgery, Cedars Sinai Healthcare System, Los Angles, CA 90048

**Corresponding Authors:**

Hamid Ghaednia, Ph.D.

hamid.ghaednia@cshs.org

Department of Orthopaedic Surgery and Department of Computational Biomedicine

Cedar Sinai Health System

Joseph Schwab, MD.

joseph.schwab@cshs.org

Department of Orthopaedic Surgery and Department of Computational Biomedicine

Cedar Sinai Health System

**Abstract**
Plantar pressure measurements can provide valuable insight into various health characteristics in patients. In this study, we describe different plantar pressure devices available on the market and their clinical relevance. Current devices are either platform-based or wearable and consist of a variety of sensor technologies: resistive, capacitive, piezoelectric, and optical. The measurements collected from any of these sensors can be utilized for a range of clinical applications including patients with diabetes, trauma, deformity and cerebral palsy, stroke, cervical myelopathy, ankle instability, sports injuries, and Parkinson's disease. However, the proper technology should be selected based on the clinical need and the type of tests being performed on the device. In this review we provide the reader with a simple overview of the existing technologies their advantages and disadvantages and provide application examples for each. Moreover, we suggest new areas in orthopaedic that plantar pressure mapping technology can be utilized for increased quality of care.

## 1. Introduction

Plantar pressure measurement, or pedobarography, quantifies or qualifies the interaction of feet with surfaces and objects. These measurements can yield insights including measures of balance and regions of peak pressures, all of which have proven valuable in a range of clinical applications, from assessing risk of diabetic foot ulceration to monitoring the progression of musculoskeletal or neurological disorders.[1,36-37] However, the clinical applications of these measurements in everyday healthcare systems is limited relative to their potential. This may be due to the considerable variation in the methods of plantar pressure measurement and in the design of the devices used to carry out such measurements. This, combined with a lack of clinical validation, has made it difficult for physicians to identify the appropriate device for their specific use cases.

Plantar pressure is typically measured either with a wearable device—this may be a sock or shoe insole—or a platform-based device. It must be considered that pressure mapping devices are significantly different from force plate platforms. Force plates most often consist of only four force sensors that measure the total applied force on a plate and the center of force. In contrast, pressure mapping devices measure the distribution of force under the foot and can provide significantly more in-depth data. The underlying sensor technologies used in plantar pressure mapping devices vary, as well as the specifications and design of each device. Sensor types used in plantar pressure measurement have developed over the past decades to include resistive sensors, capacitive sensors, piezoelectric sensors, and optical sensing modalities such as fiber Bragg gratings and frustrated total internal reflection.[2,3] These sensor technologies have been incorporated into many devices, both in-shoe and platform-based, and used across a variety of research and clinical applications, with a range of products having come to market over the past few decades.

The applications of these technologies can vary widely depending on the clinical conditions or research questions being assessed. For example, in addition to establishing ulceration risk in diabetic patients by measuring peak pressure zones, plantar pressure measurement has also been used to assess successful recovery in lower extremity injuries by examining weight bearing on the healing limb,[4] assess the impact of deformity correction surgery on balance in cerebral palsy patients,[5] as well as to assess balance issues in cervical myelopathy patients in comparison to healthy volunteers.[6]

The purpose of this review is to familiarize the reader with available plantar pressure measurement modalities and their underlying technologies, and then to describe the variety of ways in which plantar pressure has been measured to provide insight in clinical applications. Given the broad range of plantar pressure measurement applications, the scope of this paper will be limited to studies of human plantar pressure measurement for clinical applications.

## 2. Plantar Pressure Measurement (Technology)

Many devices available today are built on technological advancements made in device design and sensor technology over the past several decades.[7,8] Understanding the differences between different types of plantar pressure measurement devices is critical for determining the most appropriate method for a particular application.

### *2.1. Measurement device types:*

Plantar pressure measurement devices typically fall under one of two primary categories: platform devices and wearable devices.

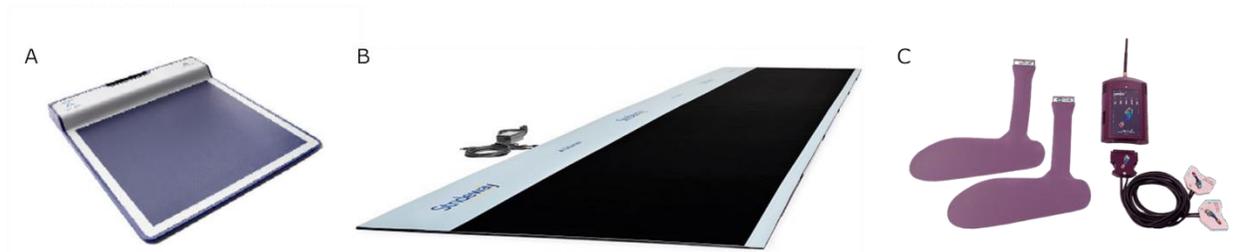

**Figure 1.** A) A platform-based device;[9] B) Multiple-platform device;[10] C) A wearable insole device.[11]

**Platform devices:** Platform- or plate-based designs provide a surface on which a patient may stand, walk, and perform activities, while pressure measurements are made from the patient's contact with the surface of the platform, Fig. 1(A).[12] These platforms may be constructed with a variety of sizes and shapes, although they typically provide enough surface area for one or two feet to rest on the platform. At the largest end of the spectrum, the Tekscan Strideway (Norwood, MA) consists of multiple platforms that can be connected to create a mat of up to 38.4 ft$^2$, Fig. 1(B).

**Wearable devices:** Wearable devices incorporate pressure measurement technology into a wearable unit, ranging from socks to shoe insoles, Fig. 1(C).[12,13] These devices remain in contact with the wearer, and measure pressure at the soles of the feet when the feet and device come into contact with an external surface or object. Such devices allow for targeted and deliberate placement of sensors at key locations along the surface of the foot, although some device designs incorporate a more evenly distributed array of sensors across a larger span of the sole. Wearable devices enable plantar pressure measurements while performing dynamic movements such as walking, running, and jumping.

*2.2. Sensor technologies*

| Sensor Type | Advantages | Disadvantages |
| --- | --- | --- |
| **Piezoresistive** | Easily constructed<br>Low cost of production with high durability<br>Resistance to dynamic pressure changes, vibration, and shock | The sensor must be powered<br>Limitations on scaling down due to sensitivity reduction and increased power consumption |
| **Capacitive** | Mechanically simple and sturdy<br>Features low hysteresis and a decent repeatability of measurements | Sensitivity to vibration<br>Certain material constraints can limit applications |
| **Piezoelectric** | Small and sensitive<br>Requires only a slight deformation to generate an output<br>Very robust and appropriate for use in a variety of harsh environs | A charge amplifier is obligatory to translate the high impedance charge output into a voltage signal<br>Only suitable for dynamic pressure measurement |
| **Fiber Bragg grating (optical)** | Light weight<br>Immunity to electromagnetic interference | Potentially fragile instruments<br>Thermally sensitive |
| **Frustrated total internal reflection (optical)** | High sensitivity<br>Immunity to electromagnetic interference<br>Low cost of production<br>Easy to scale up | Devices are thick to focal point of cameras<br>The analytical solution of constructing pressure maps from light intensity values is still not complete |

**Table 1. Examples of various pressure sensor technologies with associated advantages/disadvantages.**

**Piezoresistive sensors** detect changes in pressure through the associated changes in electrical resistance of a strain gauge fixed to a diaphragm that moves in relation to exerted pressure. Alternatively, resistive sensors may also be made of metal sensing elements whose resistance changes with deformation caused by changing pressure. **Capacitive sensors** function by detecting changes in electrical capacitance caused by the movement of an internal sensing diaphragm between two capacitor plates. **Piezoelectric Sensors** are made up of solid materials such as quartz that generate electrical output in response and proportion to slight deformations.

In addition to these traditional modes of pressure transduction, another category of pressure sensing technology is that of optical methods of collecting continuous pressure distribution data. **Fiber Bragg grating sensors** are a type of sensor that consists of an optical fiber with a series of gratings packed inside.[3] The reflected wavelengths of light can be assessed to determine the strain on the sensors. **Frustrated total internal reflection (FTIR) based sensors** measure pressure by the pattern of scattered light that radiates from points of contact between the foot and a glass plate within which light is trapped through total internal reflection.[14] Further explanation of these sensor technologies may be found in Appendix A.

*2.3. Derived metrics from plantar pressure measurements*

While plantar pressure measurement devices collect time-series data of pressure distribution maps, this raw data alone requires further processing or analysis to yield useful insights. Such processing can produce measures of balance, such as the center of contact area, center of force, or weight-bearing ratio, as well as measures showing peak plantar pressures and foot shape, Fig. 2. The data that can be extracted depends on the pressure measurement technology, as some technologies cannot detect both contact area and forces. *It must be considered that pressure mapping devices are significantly different and more advanced compared to force plates, which can only measure center of force and the total applied force in time.* Force plates fail to capture the dynamic loading of each foot and thus any asymmetrical or atypical loading patterns. Figure 2 shows a comprehensive analysis of the quantitative and qualitative measurements that can be derived from pressure sensing technologies. Each frame in the time-series data is a pressure map containing information about the centers of contact area and force, the peak pressures underneath the foot, the total contact area of the foot against the pressure-sensing platform, and the ratio of force exerted under the left and right foot, Fig. 2(A). A quantitative analysis of the measures over time can provide information about subjects' balance including sway parameters such as center of contact area and force spread and standard deviation, Fig. 2(B). The plantar pressures and contact areas can also be visualized for a qualitative analysis of foot and arch shape, foot symmetry, and peak pressures Fig. 2(C). The derived metrics provide objective insights into patient balance, foot deformities, and ulcer formations, which are useful for a variety of clinical applications.

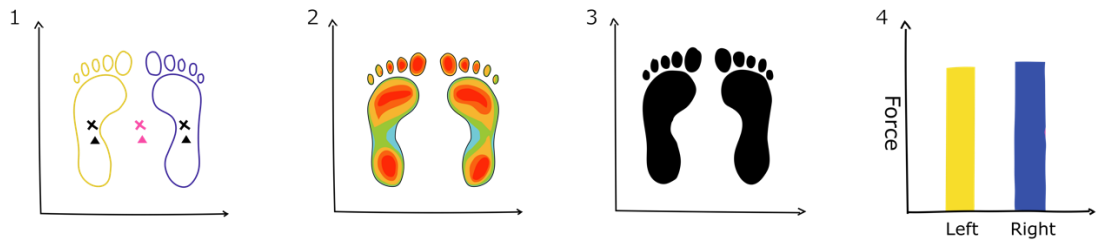

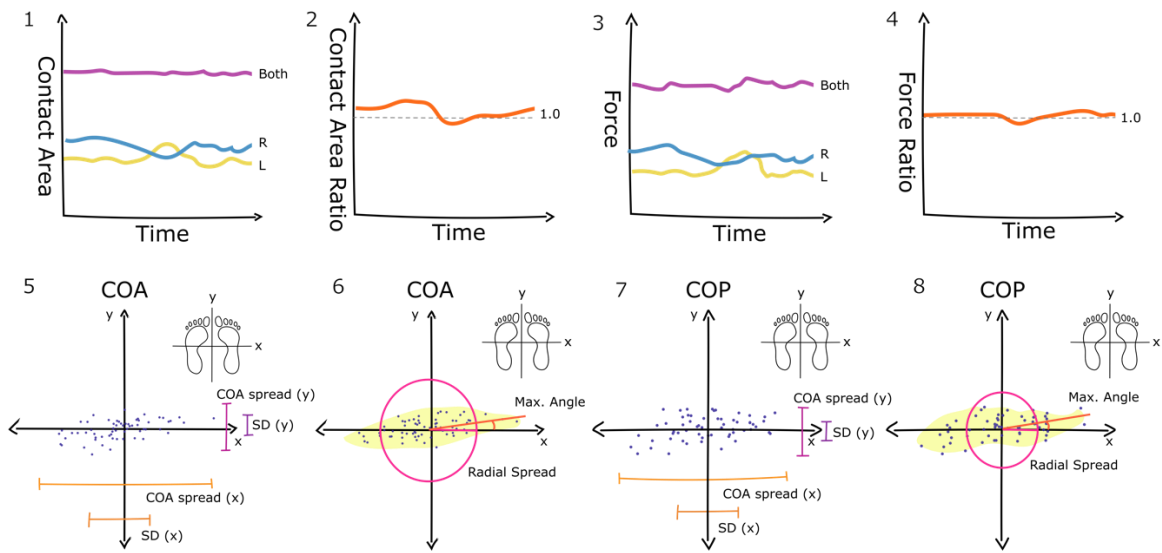

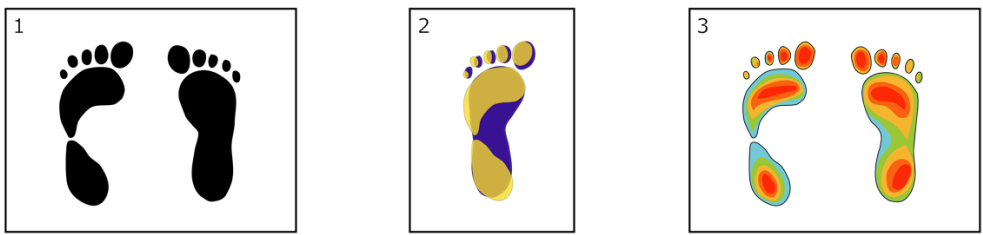

**Figure 2. Quantitative and qualitative measurements derived from plantar pressure mapping.** In a recording of plantar pressure over a test period, several analyses can be subsequently performed. A1) In each frame, the center of pressure (COP) and center of area (COA) can be calculated for the individual feet and for both feet. A2) Thermal heat maps of pressure can be constructed from optical data to show specific high- and low-pressure regions. A3) A contact area trace can be extracted to observe the shape of the contact surface and contribute to temporal calculations of change in area. A4) In each frame, the force ratio between the left and right feet can be compared. B1&2) Over the test period, the contact area under each foot and both feet can be extracted, and the ratio temporally compared. B3&4) Similar to the contact area, over the test period, the force exerted by each foot and both feet can be extracted, and the ratio temporally compared. B5&6) The centers of area of each frame (A1) are plotted with the mean at the origin. From there, the spread and standard deviation of COA shift on the sagittal, coronal, and radial directions are calculated. Additionally, performing principal component analysis (PCA) could determine the maximum angle of COA shift. B7&8) Similar to the COA, the centers of pressure of each frame (A1) are plotted with the mean at the origin. The spread and standard deviation of COP shift in the sagittal, coronal, and radial directions can be calculated. PCA can determine the angle of maximum COP shift. C1) Contact area traces can show high and low arches and other foot shapes. C2) Foot symmetry can be compared by overlaying the reflection of one foot's contact area trace over the other foot. C3) Exact points of peak pressure under the foot can be shown on a thermal pressure map.

### *2.4. Commercially Available Devices*

A set of commercially available plantar pressure measurement devices have been investigated for use in a variety of clinical contexts.

Tekscan (Norwood, MA) has an array of medical and dental pressure mapping systems that use resistive sensors in a film-based pressure system. Between in-shoe and platform-based evaluation systems, Tekscan is a leader in leveraging sampling frequency and resolution for sports medicine and other clinical evaluation. Their main product, Matscan, comes in both a single platform and walkway configuration. Hellstrant et al. utilized their in-shoe product, F-Scan, in a clinical study to evaluate peak pressure for different types of insoles and their impact on pressure distribution over a period of several months.[15] The study, consisting of 114 patients with diabetes, found that custom-made insoles in stable walking shoes reduced pressures underneath the heel. However, differences in other regions of interest were unable to be determined due to variations within the measurements.

Moticon (Munich, DE) is an in-shoe system and features a wireless, slim insole. The limitations of their system are the low sampling frequency and sensor density, but the simplicity and mobility of their technology has been a boon to their growth. In a study by Kraus et al., following Moticon's clinical use, the wearable insoles and machine learning algorithms were utilized to evaluate physical frailty as defined by the Short Physical Performance Battery (SPPB).[16] The authors found that gait analysis performed better than the physical performance tests, the timed up-and-go and the SARC-F test, when identifying physical frailty in geriatric patients.

Novel (Munich, DE) has a range of products, with both walkway and platform options. The company provides clinically relevant data, and a distinguishing factor of the hardware is their usage in diabetic foot ulceration. Becker et al. took advantage of the Novel EMED platform to investigate if surgical intervention in ankle fractures led to gait symmetry in relation to different fracture types.[17] The plantar pressure distributions showed increased loading in the forefoot of the injured leg in patients with satisfactory gait symmetry and decreased loading under the metatarsal heads of patients with poor symmetry. The study found no relationship between asymmetries and the fracture type or clinical outcome.

Across these technologies, a variety of sensor types are utilized to generate imaging of the plantar surface. In addition, the technological specifications of the devices are based on different needs of the user, hence, pressure ranges, sampling frequency, and device cost are variable. As such, Table 2 features a group of devices and their associated characteristics providing a synopsis of commercially available devices and their specifications.

| | Product Information | | | Specifications | | |
|---|---|---|---|---|---|---|
| **Device** | **Sensor Technology** | **Device Type** | **Manufacturer** | **Frequency** | **Range** | **Cost** |
| F Scan | Resistive | In-shoe | Tekscan | 100 Hz | 0-862 kPa | $12k |
| MatScan | Resistive | Platform | Tekscan | 100 Hz | 345-862 kPa | $6k |
| HR Mat | Resistive | Platform | Tekscan | 185 Hz | 345-1103 kPa | $13k |
| Strideway | Resistive | Platform | Tekscan | 500 Hz | 276-862 kPa | $23k-$100k |
| Mobile Mat | Resistive | Platform | Tekscan | 100 Hz | 345-862 kPa | $7k |
| emed | Capacitive | Platform | Novel | 50 Hz | 10-1270 kPa | $37k |
| pedar | Capacitive | In-shoe | Novel | 100 Hz | 15-600 kPa | $28k |
| Moticon | Capacitive | In-shoe | Moticon | 100 Hz | 0-500 kPa | $0.6k |

**Table 2. An overview of current technology and their respective technical qualities.** These devices are both platform-based and wearable devices with a variety of technical specifications. Increased frequency and range correspond to significant increases in cost, up to $100K for the Tekscan Strideway.

## 3. Clinical Applications

We conducted a scoping review to identify and summarize the applications of plantar pressure mapping. The primary database utilized was Google Scholar, with additional sources extracted from the references of the originally selected papers. The keywords used to identify literature for the initial search were *plantar pressure measurement, pressure distribution, balance, falling/fall prevention, foot and ankle,* and *sports medicine*. Rayyan software was utilized to conduct blinded screening of 91 abstracts over the course of 3 weeks. After screening by 5 researchers, 44 articles were chosen for full review. As displayed in Figure 3, these articles were read in full, and relevant device technologies and clinical applications were extracted from each article and managed in Microsoft Excel.

The inclusion criteria for technical information of the studies were works that evaluated plantar pressure mapping with subheadings of force measurement and mapping. The selected clinical disorders for our literature search were diabetic ulceration, neurological instability, and orthopedic disorders.

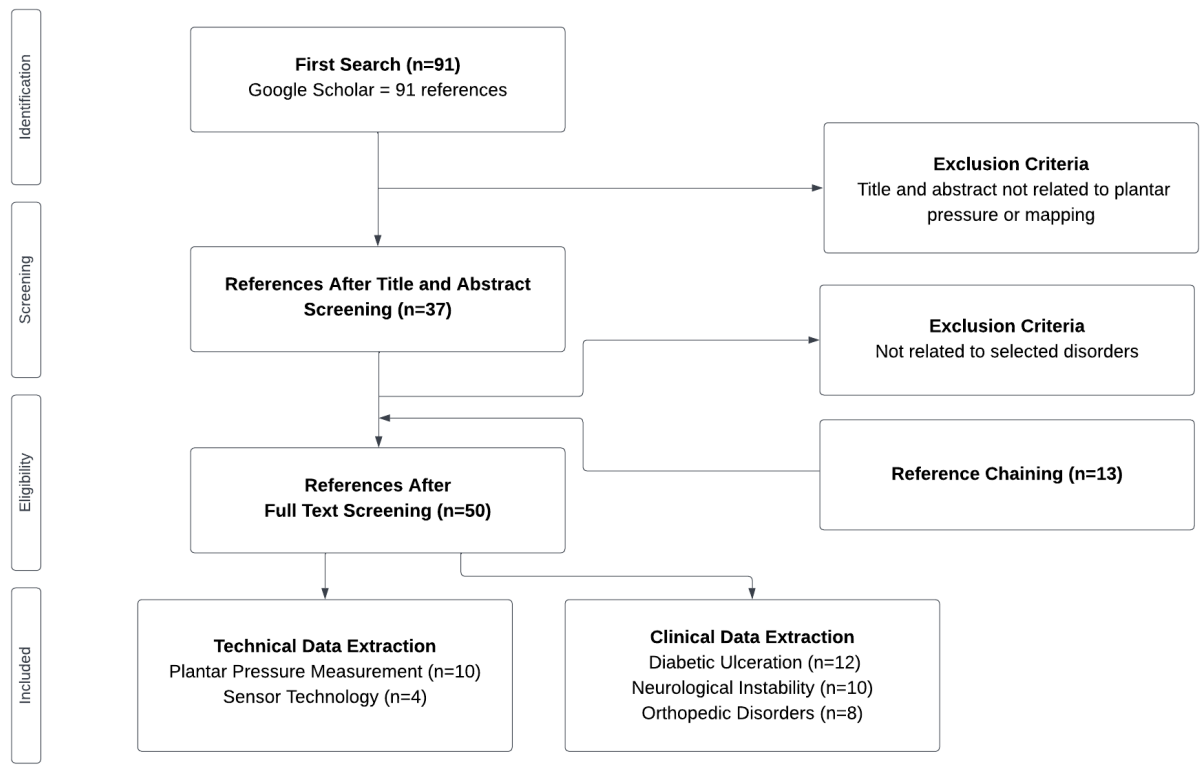

**Figure 3. Systematic review flow chart.** The 14 papers not utilized in clinical data extraction were related to the sensing and device technology used, separate from the clinical disorders examined.

### 3.1. Diabetes

Several studies have used plantar pressure measurement to assess the risk of lower extremity complications in diabetic patients, who are often at particularly high risk for problems including foot ulceration.[1] Rodgers et al. used a piezoceramic platform-based device to assess peak pressures and pressure distributions in normal patients in order to establish a baseline for comparison with patients at risk for diabetic foot ulceration.[18] Consistent with multiple additional studies, Rodgers et al. found peak plantar pressure distributions to be positively correlated with diabetic foot ulceration risk. Other studies found plantar pressure distributions to be predictive of diabetic foot ulceration risk, using either weight distribution ratios from the forefoot to the hindfoot, or peak pressure values.[16,17,19]

Orthotics designed to prevent diabetic foot ulcers and other common interventions are commonly studied using pressure platforms. In a two-year randomized trial on 114 patients with type 1 or type 2 diabetes, Hellstrand et al. showed that peak pressure in the heel of patients wearing custom-made insoles was lower than for mass-manufactured insoles. The estimated difference in peak pressure between the custom insoles and the prefabricated insoles was 63 kPa with a p-value less than 0.001. The study compared peak plantar pressure and pressure time integral measured using an insole measurement system, enabling the authors to conclude that custom-made insoles were effective in reducing pressure levels at the heel.[15] Similar studies have been carried out assessing the effectiveness of insoles, padding, dressing or socks, and other orthotics in reducing peak plantar pressures in diabetic patients at risk of ulceration.[20–24]

Plantar pressure measurement assessment can additionally be implemented to customize orthotics for diabetic patients. Guldemond et al, assessed the effect of different insole configurations on the peak plantar pressure of the patients with diabetic foot using an in-shoe pressure system.[25] They concluded that the combination of extra arch support (10mm) and metatarsal dome had the largest decreasing effect on the peak plantar pressures in the central and medial forefoot. Preece et al, investigated how different rocker sole parameters may affect plantar pressures of diabetic patients.[26] Overall, the design of the insoles can be based on the pressure changes and also to offload areas that are prone to wounds and ulcers.

Factors such as leg lengths discrepancies in patients with diabetes have also been investigated utilizing plantar pressure mapping systems. El-nahas et al. examined the effect of a simulated leg length discrepancy on pressure distributions in 28 diabetic patients with neuropathic foot ulceration.[27] The peak pressure significantly increased for the shorter leg, and plantar pressure mapping was able to detect that diabetic patients with leg length discrepancies may be at greater risk of diabetic foot ulceration given the likelihood of higher peak plantar pressures. Plantar pressure mapping enables the assessment of the risk of lower extremity complications, such as foot ulceration, and the investigation of the effects of varying body anatomy in diabetic patients.

*3.2. Trauma*

Fracture healing is another common area of investigation in which plantar pressure measurements have been applied. Davies et al. examined the outcomes of open reduction and internal fixation of displaced os calcis fractures using a dynamic optical platform-based device for 12 patients. The high spatial resolution of the optical device allowed the authors to conclude that some plantar functions, such as subtalar and calcaneocuboid joint function, are not remedied through surgical intervention.[28] In the work done by Neaga et al., patients were monitored during post-traumatic rehabilitation with an in-shoe device incorporating pressure transducers to alert patients of excessive loading of limbs.[29] Contreras et al. assessed healing in patients after internal fixation of calcaneal fractures using an in-shoe device to measure differences between contact area, pressure, and strength in the forefoot and rearfoots.[30] In this work, plantar pressure mapping enabled the authors to verify no significant change between the mean peak pressure in the hindfoot and forefoot between the operated (late postoperative) and normal side. Additionally, Becker et al. measured gait symmetry using platform-based system in patients who had undergone surgical treatment of ankle fractures. The plantar pressure analysis showed significant load asymmetries, illustrating compensation mechanisms that are used to regain gait symmetry after ankle alteration by trauma.[16] In these works, researchers used plantar pressure measurements to assess fracture rehabilitation progress and investigate the effects of surgical treatments.

Plantar pressure distributions have also been used in measuring sports-related trauma injury outcomes and investigation into prevention methods. A study by Tatar et al. used in-shoe plantar pressure measurements to measure load distribution on the foot and extremities of amputee football players to determine the points of greatest load bearing during sport activity.[31] In the study, in-shoe devices and sensors fixed to the subjects' gloves monitored loads while 15 amputee football players walked, ran, and kicked the ball. The authors found that the kicking movement greatly increased the loading and transferred most of the load to the hands. Another study conducted by Guy-Cherry et al. assessed different landing styles for their risk of injury using the same in-shoe plantar pressure measurement system. The pressure measuring system was used to measure the ground reaction forces under subjects' feet while landing in different styles: stiff, soft, and self-selected. Subjects were found to have the highest peak pressure values when landing with the soft style, confirming the authors hypothesis that landing with a greater knee flexion decreases peak ground reaction forces and thus load on the anterior cruciate ligament.[32] Morrison et al. utilized plantar pressure mapping in combination with rearfoot alignment measures in assessing the points of greatest plantar loads in walking motions of athletes with plantar fasciitis.[33] The authors found that subjects with chronic ankle instability have a more lateral foot positioning and loading pattern during a barefoot running gait compared to controls. In these studies, plantar pressure measurement was an essential tool to gain a quantitative measure of patients' condition in sport related trauma injuries and prevention.

### 3.3. Deformity and Cerebral Palsy

Plantar pressure measurement is also used to assess balance and pressure distribution in patients with cerebral palsy. Leunkeu et al. examined the differences in plantar pressures between children with cerebral palsy and children without cerebral palsy during normal walking. Subjects with cerebral palsy were found to have higher peak pressures under the lateral and medial columns of the foot as compared to able-bodied subjects.[34] In a study of 40 children with Congenital Talipes Equinovarus (CTEV), Salazar-Torres et al. used a platform pressure analysis system to determine any differences in pedobarographic outcomes between those treated with the Ponseti technique and those treated with a more traditional approach. Through investigation of average and maximum peak pressures, and pressure time integral of each region, the authors were able to conclude that children treated with the Ponseti technique had greater pressure under the lateral border of the mid-foot.[35] Park et al. assessed plantar pressure distribution in patients surgically treated for calcaneal deformities. The authors used multiple force platforms for gait analysis and a high-resolution pressure assessment system to measure dynamic foot pressure. Corrective surgical procedures were found to effectively reduce the pressure on the calcaneus, increase pressures in the midfoot and forefoot, and prevent recurrence of the deformity.[36] Chang et al. examined plantar pressure distributions in children wearing a gait-correcting orthosis vs. those without. A motion capture system was used alongside a pressure platform to capture gait parameters simultaneously with plantar pressures. The authors found the gait-correcting orthosis improved all gait parameters and altered the path of pressure trajectory towards the midline.[37] In addition, Duckworth et al. compared plantar pressure distributions pre- and postoperatively in children suffering from spina bifida undergoing surgical correction of equinovarus feet, finding reductions in peak values.[38]

Orthotics for treatment of pes planus have also been widely studied using plantar pressure measurement systems. In 1996, Miller et al, conducted a trial to examine how rearfoot orthotics (medial heel wedge insoles) may affect ground force reactions in asymptomatic pes planus individuals.[39] They reported a reduction in vertical and anteroposterior ground force reactions in the early stages of stance phase of the gait cycle. Tang et al, assessed the change in plantar pressures of patients with flatfoot using a custom made, total contact insole, with an extended heel guard and forefoot medial posting.[40] The authors used a plantar pressure mapping system to demonstrate that wearing the combination of insoles and sports shoes significantly decreased peak pressures in the hallux and heel areas when compared to wearing only sports shoes. In these studies, with plantar pressure mapping systems, the authors were able to assess plantar pressures and use that to inform the effectiveness of treatments such as insoles. Plantar pressure measurement systems have also been used to design custom insoles based on patient-specific deformities. As an example, Jiang et al. used a plantar pressure mapping platform to conduct detailed plantar characteristic analysis on patients with flatfoot then designed an insole to redistribute the pressure.[41] Their custom insoles were found to improve the plantar pressure distribution and gait efficiency of patients with flatfoot.

### 3.4. CNS Diseases

Mkorombindo et al. leveraged plantar pressure measurements with the use of a platform-based plantar pressure mapping system utilizing capacitive sensors to assess balance in cervical myelopathy patients.[6] The study quantified standing balance during the Romberg Test using center of pressure movement, sway area, and other parameters. They found that these parameters could quantify poor standing balance in patients with cervical spondylomyelopathy against age-matched healthy volunteers.

Shalin et al. trained a convolutional neural network on in-shoe plantar pressure measurements to predict freezing of gait in patients with Parkinson's disease. In this case, an in-shoe device enabled natural walking trials of longer durations, providing more data for training predictive models. Models trained solely on foot pressure distributions had high sensitivities and specificities, demonstrating the ability of pressure distributions to predict freezing of gait.[42]

Studies conducted by Kim et al. and Bayouk et al. have assessed the efficacy of balance training in recovering stroke patients, using plantar pressure measurement-derived assessments of balance to evaluate improvements in balance and symmetric posture.[43,44] Kim et al. found the plantar pressure measurement-derived assessments (center of pressure, peak pressure in the hindfoot, and hindfoot contact area) differed significantly between the control and repetitive sit-to-stand training group.[43] Bayouk et al. examined the effect of a task-oriented exercise program on balance parameters, including center of pressure variability and total excursion, finding that the program with sensory manipulation was more effective at improving balance than a conventional program.[44] Another study by Carver et al. examined the effect of physical exertion on balance in stroke patients, finding that physical exertion can increase imbalance and instability in hemiparetic stroke patients.[45]

## 4. Discussion

Plantar pressure measurement has played an essential role in a wide range of clinical applications. Over the past several decades, a variety of pressure measurement approaches have been applied to yield insights and contribute to clinical care in many areas of medicine, ranging from diabetic foot ulceration to cervical myelopathy.[1,6]

With any given application, it is important to consider the specific role played by pressure measurement. While each analysis is likely to be unique in some way, the priorities and preferences for a particular use of plantar pressure measurement may highlight certain approaches over others. For instance, in a study assessing standing balance in patients with cervical myelopathy, a platform type device may be suitable.[6] On the other hand, in a study tracking the recovery progress of athletes following surgery for bony or soft tissue injury, it might be important for a device to measure dynamic pressures during walking or other measured activities that go beyond the spatial confines of a typical platform device.[46] In this case, an in-shoe device may present advantages. The tradeoffs, if any, may not be simple to navigate, and in-shoe devices can present limitations in measurement performance, tethering of the wearer's soles to a separate sensing apparatus, and perhaps changing the weight distribution or subject behavior. A further consideration beyond methodological tradeoffs and priorities is that of cost, which may render some options more feasible or appealing than others.

It is important to consider the differences between pressure measurement methods (Table 1) as there are a variety of plantar pressure measurement products available today. An examination of five common pressure measurement devices on the market in 2010 found high static and dynamic accuracy for both capacitive and resistive technologies.[47] Piezoelectric and optical sensors are less common due to their relatively recent development into plantar pressure sensing applications, but have the advantage of increased sensitivity.

In addition, there is great variation in the costs associated with different devices and methods of pressure measurement, and while technological development has been accompanied by some reduction in costs over the decade, the choice of measurement device may often be constrained by flexibility of funding.

Although clinical applications of plantar pressure measurement have been broad and varied thus far, the range of applications may continue to expand further as device technology improves and as our understanding of plantar pressure distributions as they relate to health and a variety of medical conditions continues to develop. Improvements in accessibility may allow for such devices to become a part of routine clinic visits, and advancements in technology and design may permit everyday usage of wearable devices. Such developments would allow for more frequent collection of data on general physical health and daily movements and activities, which may provide valuable insight to patients and clinicians alike.

Many studies have used each mode of plantar pressure measurement, whether an in-shoe wearable device, a platform-based device, or some combination thereof. Despite the potential advantages of in-shoe devices in portability and in ability to replicate in-shoe dynamics, they may lag behind platform-based devices in precision and accuracy. In-shoe devices, depending on the model, may also require tethering or be constrained by battery life considerations that can limit their utility for extended periods of time and their applicability in environments would otherwise be possible with a portable and wearable device.

Advancements in technology have been accompanied by a broadening of clinical applications of pressure measurement devices. As newer technologies become available and improvements are made to existing device designs, the opportunities for clinical applications of plantar pressure measurement will continue to expand—and it will remain important for investigators to understand the range of available methods of pressure measurement best suited for their purposes.

## 5. Conclusion

Plantar pressure measurement has played a crucial role in a wide range of clinical applications. As the underlying technology and device design of plantar pressure measurement continues to evolve, the opportunities for clinical applications of these devices may grow in proportion. Investigators are encouraged to consider whether such technology may be used to support or enable analyses in novel ways. Plantar pressure mapping technology yields a multitude of parameters, most of which require further investigation. Center of pressure, center of contact areas, peak pressures, and total contact area can be analyzed to provide metrics such as body sway, foot symmetry, left/right weight symmetry and more. Beyond the applications described in this work, plantar pressures could be further utilized for progress assessment, quantification of functional outcomes after surgery, quantification of risk of fall, and early detection of pathologies including dementia. Regardless of the application, it is important to carefully consider the particular needs and priorities for the project at hand when deciding to pursue one approach of pressure measurement over another.

## Appendix A
### A1. Capacitive Sensors

Capacitive sensors consist of two capacitor plates containing an electric charge which have an elastic dielectric material between them. Under pressure, the dielectric material bends and yields a voltage change that is measurable and corresponds proportionally to the amount of pressure on the sensor. The capacitance between two parallel plates of area $A$ separated by distance $x$ is $C = \varepsilon_0 \varepsilon_r \left(\frac{A}{x}\right)$, wherein $\varepsilon_0$ is the dielectric constant of free space and $\varepsilon_r$ is the relative dielectric constant of the insulator.[48] Fundamentally, it is feasible to monitor displacement by altering $\varepsilon_r$, $A$, or $x$. The method that is most utilized involves manipulating the separation between plates. Modulating the distance between plates will generate a variable metric of capacitance.[49] One of the pressure mapping device that utilizes this type of sensor is the emed® pressure platform created by Novel Co. (Novel, Germany).[11]

### A2. Resistive Sensors

There are several types of resistive pressure sensors, namely potentiometers, strain gauges, and force-sensing resistors (FSRs).

Potentiometers can measure translational or rotational displacement within a certain range, with resistive elements excited by either AC or DC voltage. As the sensor is compressed and the length of the resistive element shortens, the total resistance decreases. At low resistance ranges, stepless dynamic measurements can be made using a straight wire, but larger resistances can only produce stepwise measurements due to a helical, wire coil construction. Herein lies the inherent limitation of potentiometric resistive sensors, as the resolution is limited by the construction of the sensor itself.[48]

Strain gauges measure pressure on a nanoscale based on properties of the wire itself. A wire with length $L$, cross-sectional area $A$, and resistivity $\rho$ has an inherent resistance of $R = \frac{\rho L}{A}$. Stretching or compressing the wire length-wise changes both $L$ and $A$, resulting in a change in resistance and thus measured pressure. For some highly ionic bonded materials (typically semiconductors or conductive polymers), there are domains of electric dipole moments that are usually randomly oriented but can align in specific directions to add a piezoresistive effect, given as $\frac{\Delta \rho}{\rho}$, to either generate a mechanical (motor) or electrical (generator) effect. This occurs because the ionic clusters are brought closer together during material compression so that either conduction or quantum tunneling can occur. In the pressure sensor, pressure creates a voltage due to the generator effect that can be measured to quantify the applied force.[48]

Force-sensing resistors (FSRs) function with a nearly identical mechanism to piezoresistive strain gauges; however, the backing of a strain gauge sensor deforms with applied force, while an FSR backing does not deform. The MatScan® and F-Scan® systems by Tekscan both utilize FSR technology, while the FlexiForce® system by Tekscan and Tactilus® by Sensor Products Co. utilize piezoresistive sensors.[50]

### A3. Piezoelectric Sensors

Another type of sensing technology is piezoelectric sensors. These sensors output a voltage in response to pressure. Electric fields are produced by piezoelectric materials due to mechanical strain; conversely, an electric potential may yield physical deformation of a material.[48] These sensors have high impedance and are susceptible to electrical interference that leads to unacceptable signal-to-noise ratio (SNR). The piezoelectric effect is found in non-conducting materials such as crystals (quartz), ceramics, and thin flexible (polyvinyl lidenefluoride) PVDF films. PVDF and polymer-based sensors can be made thin, flexible and deformable. A thin metallic layer is applied to both sides to collect electrical charges and allow for connections. Piezoceramic materials are temperature sensitive and should be kept in controlled ambient conditions. Except for PVDF sensors, piezoelectric sensors are highly elastic, show little deformation and exhibit low hysteresis. As a result, they are suitable for high-frequency events.[48]

### A4. Fiber Bragg grating Sensors

Optical sensors take advantage of various properties of light. Fiber bragg Grating (FBF)/Polymer Optical Fiber (POF) sensors reflects certain wavelengths and transmits others. The wavelength that is reflected is dependent on the strain placed on the sensor. Therefore, if the reflected wavelength is measured, the strain and therefore the pressure on the sensor can be calculated. The change in wavelength relative to the strain is defined by: $\frac{\Delta\lambda}{\lambda} = C_S \epsilon + C_T \Delta T$. $C_s$ is a coefficient of strain, and $C_T \Delta T$ is a temperature term that is small enough to be neglected when using FBG as a pressure sensor.[51]

### A5. Frustrated Total Internal Reflection Sensors

The FTIR focuses on the intensity of reflected light instead of the wavelength. When light is trapped in a transparent medium, such as glass, an object that comes within that light's wavelength to the glass will reflect back light. A camera located under the transparent medium can record this reflected light. The images are then analyzed, and the light intensity in each region corresponds to the pressure.[14]